\documentclass[preprint2]{aastex}

\shorttitle{Stereoscopic electron spectroscopy}

\shortauthors{Kontar and Brown}

\begin{document}


\title{Stereoscopic electron spectroscopy of solar
hard X-ray flares with a single spacecraft}


\author{Eduard P. Kontar and John C. Brown}
\affil{Department of Physics and Astronomy, The University of
Glasgow, G12 8QQ, UK}

\email{eduard,john@astro.gla.ac.uk}



\begin{abstract}
Hard X-ray (HXR) spectroscopy is the most direct method of
diagnosing energetic electrons in solar flares. Here we present a
technique which allows us to use a single HXR spectrum to
determine an effectively stereoscopic electron energy
distribution. Considering the Sun's surface to act as a 'Compton
mirror' allows us to look at emitting electrons also from behind
the source, providing vital information on downward-propagating
particles. Using this technique we determine simultaneously the
electron spectra of downward and upward directed electrons for two
solar flares observed by the Ramaty High Energy Solar
Spectroscopic Imager (RHESSI). The results reveal surprisingly
near-isotropic electron distributions, which contrast strongly
with the expectations from the standard model which invokes strong
downward beaming, including collisional thick-target model.
\end{abstract}



\section{Introduction}

Energetic electrons and ions have long been considered (Ellison
and Hoyle, 1947) as possibly playing a key role in energy release
by magnetic reconnection, in phenomena such as solar flares
(Aschwanden, 2002), which are archetypes of magnetic explosions
and particle acceleration in the cosmos, as well as being central
to space weather and its terrestrial influence.

The spectrum of energetic electrons and the anisotropy of source
electrons are normally inferred separately. Hard X-ray (HXR)
directivity itself has, until now, been measured using two
different techniques. The first is direct simultaneous measurement
of flux spectra from at least two spacecraft at well separated
locations (Li et al, 1994; Kane et al 1988). The second method is
based on the statistical study of the distribution of HXR fluxes
and spectral indices over heliocentric angle (Datlowe et al. 1977;
Bogovalov et al. 1985; Vestrand 1987). However, neither of these
methods provides direct information about downward emitted photons
which are crucial in model testing.

High resolution HXR spectrometry with RHESSI enables
reconstruction of volume-averaged source electron flux spectra
(electrons cm$^{-2}$ s$^{-1}$ keV$^{-1}$ ) from bremsstrahlung HXR
emission spectra (photons electrons cm$^{-2}$ s$^{-1}$ keV$^{-1}$)
observed at the Earth (Piana et al. 2003; Kontar et al. 2004).
However, the observed spectrum is contaminated by an albedo
component, due to Compton back-scattering ('reflectivity') in the
dense photosphere, of those primary bremsstrahlung photons which
were emitted downward (Tomblin, 1972; Bai and Ramaty, 1978). Until
now this albedo contribution has been regarded as a nuisance to be
corrected for, either following Bai and Ramaty (Johns and Lin
1992, Alexander and Brown 2002;), or self-consistently (Kontar et
al, 2006), before spectral inference of electron spectra.

In this Letter we show that, on the contrary, the albedo spectral
'contaminant' in fact offers very valuable insight into the
anisotropy of the flare fast electron distribution. It does so by
providing a view of the HXR flare from behind, like a dentist's
mirror, except that the solar albedo mirror is spectrally
distorting so its contribution to the overall spectrum can be
distinguished. This reflectivity has a broad spectral peak in the
30-50 keV range (Bai and Ramaty, 1978; Kontar et al, 2006). It
decreases fast at low energies due to photoelectric absorption
while high energy photons are more likely to be lost to an
observer because they penetrate too deep into the solar atmosphere
to be scattered back to an external observer. On the other hand,
an important property of bremsstrahlung emission is that it is
always monotonically decreasing for any electron spectrum,
typically having a roughly power-law shape of spectral index which
is nowhere less than 1 (Koch and Motz, 1959). Thus, the primary
emission and the albedo 'bump' spectral signatures are very
distinct. The observed spectrum in the observer's direction should
contain an albedo bump feature, the strength of which is an
indicator of the degree of downward beaming of the electron
distribution. By use of this solar 'mirror' we can achieve a
degree of 2-directional beam electron beam spectrometry from
single spacecraft photon spectrometry.

We show furthermore that this insight constrains the directivity
of flare electrons, so strongly that the conventional solar flare
models with downward beaming are excluded. Note that the inferred
electron spectra and distributions cannot be explained by the
collisional scattering (Trubnikov 1965) as assumed in collisional
thick-target (CTT) model (e.g. Brown 1971; Brown 1972). In the CTT
scenario, even allowing for collisional scattering, the
downward-propagating electrons emit HXR bremsstrahlung which is
also substantially downward-collimated.


\section{Photospheric albedo as a natural constraint
on downward beaming of electrons}

For an inhomogeneous bremsstrahlung source of volume $V$, plasma
density $n({\bf r})$, and electron flux spectrum $F(E,{\bf
\Omega'}, {\bf r})$ per unit solid angle in direction ${\bf
\Omega'}$ can be averaged over the source volume. The photon flux
spectrum at Earth distance $R$ in direction ${\bf \Omega}$ is
\begin{eqnarray}
I(\epsilon) = \frac{{\bar n}V}{4\pi R^2}
\int_{\Omega'}\int_\epsilon^\infty {\bar F}(E,{\bf \Omega'})
Q({\bf \Omega},{\bf \Omega'},\epsilon, E)
\mbox{d}E\mbox{d}{\bf\Omega '} \label{emiss1}
\end{eqnarray}
where $Q({\bf \Omega},{\bf \Omega'},\epsilon,E)$ is the
bremsstrahlung cross section differential in $\epsilon$ and ${\bf
\Omega}$; $\bar n = V^{-1}\int n({\bf r}) \, dV$ is a volume
averaged density, the density-weighted mean electron flux is $\bar
F(E,{\bf \Omega'}) = (\bar n V)^{-1}\int n({\bf r}) F(E,{\bf
\Omega'},{\bf r}) \, dV$, and $\epsilon$, and $E$ are photon and
electron energies correspondingly. In fact $Q$ depends only on the
angle $\widehat{{\bf\Omega'}{\bf \Omega}}$ between the incoming
electron ${\bf\Omega'}$ and the emitted photon ${\bf\Omega}$
directions.

Equation (\ref{emiss1}) can be approximated using a two
directional representation based on the fact that here we are
concerned with emission upward ($u$) toward the observer and
downward ($d$) toward the scattering photosphere. The flux toward
the observer $I_o(\epsilon,\theta)$ from a source with
heliocentric angle $\theta$ can be written
\begin{eqnarray}
I_o(\epsilon) = {1 \over 4\pi R^2} \, \, \bar n V
\int_\epsilon^\infty \left[ Q^F(\epsilon,E){\bar
F}_u(E)+\right.\cr \left.Q^B(\epsilon,E){\bar F}_d(E)\right] dE,
\label{I_o}
\end{eqnarray}
where $\bar F_{u,d} = (\bar n V)^{-1}\int F_{u,d}(E,{\bf r}) \,
n({\bf r})dV$ and, using axial symmetry, we have introduced
\begin{eqnarray}\label{angle_av}
Q(\epsilon,E,\theta _0)=\frac{1}{\cos(\theta _0-\Delta
\theta)-\cos(\theta _0+\Delta \theta)}\cr\int _{\theta _0-\Delta
\theta}^{\theta _0+\Delta \theta}Q(\epsilon, E, \theta
^{\prime})\sin(\theta ^{\prime})d \theta ^{\prime}
\end{eqnarray}
the cross-section averaged over $[\theta _0-\Delta \theta, \theta
_0+\Delta \theta]$ and centered at angle $\theta _0$. Hence,
$Q^F(\epsilon,E)\equiv Q(\epsilon,E,\theta _0 = 0)$ and
$Q^B(\epsilon,E)\equiv Q(\epsilon,E,\theta _0 = 180^o-\theta )$,
where $\theta $ is the heliocentric angle of the source. Electron
spectrum ${\bar F}(E, \theta)$ is defined in a similar two
directional approximation: ${\bar F}_u(E)$ and ${\bar F}_d(E)$ are
the density weighted volumetric mean flux spectra of electrons
directed towards the observer 'upward' and downward respectively,
also averaged over $\Delta \theta$.

The probability of photon emission along the direction of motion
is around ten times higher than in the opposite direction and four
times that at right angle, the cross-section having a typical
angular scale $\Delta \theta \sim 30-50^o$ (Koch and Motz, 1959)
for energies 50-300 keV. Moreover, the electron distribution
should have some additional angular spread (e.g. because of
collisions (Trubnikov, 1965; Brown 1972, Leach and Petrosian 1983,
MacKinnon and Craig, 1992) and magnetic field convergence).
Therefore, for the spatially averaged distribution it is natural
to deal with angular distribution averaged over $\Delta \theta$.
In our calculations we take $\Delta \theta =45^o$ as a
characteristic scale of angular dependency. (In the limit $\Delta
\theta =180^o$ the cross-section becomes angle averaged and
equation (\ref{I_o}) is reduced to the simplified equation for
isotropic electron distributions used in, e.g., Brown et al.,
(2003)).

X-ray photons propagating downwards undergo Compton backscattering
and absorption in the dense photosphere, the spatially integrated
reflected flux being expressible via a Green's function approach
(Magdziarz and Zdziarski, 1995; Kontar et al, 2006).
\begin{equation}\label{fluxR}
    I_{r} (\epsilon) =\int _{\epsilon}^{\infty}
    G(\mu,\epsilon,\epsilon ^{\prime}) I_{d} (\epsilon ^{\prime})
    d \epsilon ^{\prime}
\end{equation}
where $I_{d} (\epsilon )$ is the the downward directed flux and
$G(\mu,\epsilon,\epsilon ^{\prime})$ is the angular ($\mu =\cos
(\theta)$) dependent Green's functions (Kontar et al, 2006). The
reflected flux can be written
\begin{eqnarray}
\label{I_r} I_r(\epsilon,\mu) = {{\bar n} V \over 4\pi R^2} \, \,
\int _{\epsilon}^{\infty}
    G(\mu,\epsilon,\epsilon ^{\prime})
    d \epsilon ^{\prime} \cr\int_{\epsilon ^{\prime}}^\infty \left[
Q^F(\epsilon ^{\prime},E){\bar F}_d(E)+\right. \left.Q^B(\epsilon
^{\prime},E){\bar F}_u(E)\right] dE,
\end{eqnarray}
where we use the same cross-sections as in Equation
(\ref{angle_av}).

The total observed flux is the sum of direct and back-scattered
X-rays, i.e., $I(\epsilon) = I_{o} (\epsilon) + I_{r} (\epsilon)$
with $I_o(\epsilon)$,  $I_r(\epsilon)$, given by Equations
(\ref{I_o}), (\ref{I_r}). Even using a single spacecraft we thus
always observe a combination of reflected downward flux and direct
flux which are functionally very different. This fact, and the
properties of bremsstrahlung emission, allow us to draw, from
spectral data, conclusions about both the directivity and the
spectrum of the emitting electron distribution.

\begin{figure}
\begin{center}
\plotone{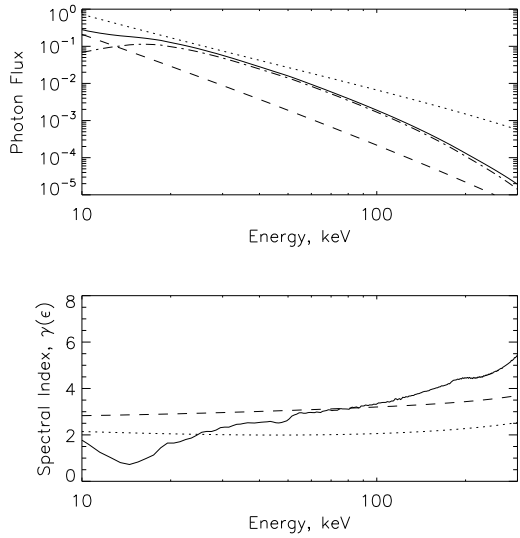}
\end{center}
\caption{Simulated photon flux spectra (upper panel) and spectral
index (lower panel) for a flare located at heliocentric angle
$\theta=40^o$ with electron distribution (${\bar F}_d(E)\sim
E^{-1.5}$ and ${\bar F}_u(E)=0$, e.g., all electrons are directed
downward and confined within pitch angle of $\pm 45^o$). Upper
panel: Observed (solid line), downward directed (dotted line),
upward directed (dash line) and reflected (dash-dotted line) flux
spectra. Lower panel: observed (solid), upward (dotted), and
downward (dashed) spectral indexes.} \label{2d_example}
\end{figure}

Figure (\ref{2d_example}) shows that albedo sets a natural
constraint on possible directivity present in solar flares. Strong
downward directivity (all electrons are confined within a pitch
angle of $45^o$) typical to the collisional models (e.g. Brown,
1972; MacKinnon and Craig, 1992) leads to flatter than observed
spectral indices below 20 keV and softer than observed spectra in
the hundreds of keV range. It can be seen that the observed
spectral index at 200-300 keV cannot be less than $5$ (Figure
(\ref{2d_example}), much softer than typical spectral indices
$2-3$ (Vestrand et al, 1987; Li et al, 1994).

\section{Two directional electron distributions}
In reality we deal with discrete data sets rather than continuous
functions. Therefore, the data vector of photons ${\bf I}(\epsilon
_i)={\bf I_o}+{\bf I_r}$ can be written
\begin{equation}\label{IP}
{\bf I}=\left(
    \begin{array}{cc}
{\bf Q^F}+{\bf G}(\mu){\bf Q^B}\;\;\; {\bf Q^B}+{\bf G}(\mu){\bf
Q^F}
\end{array}
\right) \left(
\begin{array}{c}
{\bf {\bar F}_d }  \\
{\bf {\bar F}_u }
\end{array}
\right)
\end{equation}
where $\epsilon _i$ for $i=1...N$, and $E_j$ for $j=1...M$, where
${\bf {\bar F}_{d,u}}(E_j)$ are the electron data vectors and
${\bf G}, {\bf Q}^{B,F}$ are matrix representations of the kernels
of integral equations (\ref{I_o},\ref{I_r}). Green's matrixes
${\bf G}(\mu)$ depend on heliocentric angle of the source $\mu =
\cos (\theta)$ and have been calculated in Kontar et al., 2006.
The multiple sum in Equation (\ref{IP}) $\Sigma ^M\Sigma2$ can be
expressed as a single sum $\Sigma^{2M}$ (Hubeny and Judge 1995).
Equation (\ref{IP}) can be solved using the Tikhonov (1963)
regularization method of which our implementation was successfully
tested by simulation (Kontar et al. 2004) and applied to RHESSI
data (Kontar et al. 2005).

\begin{figure}
\begin{center}
\epsscale{1.0} \plotone{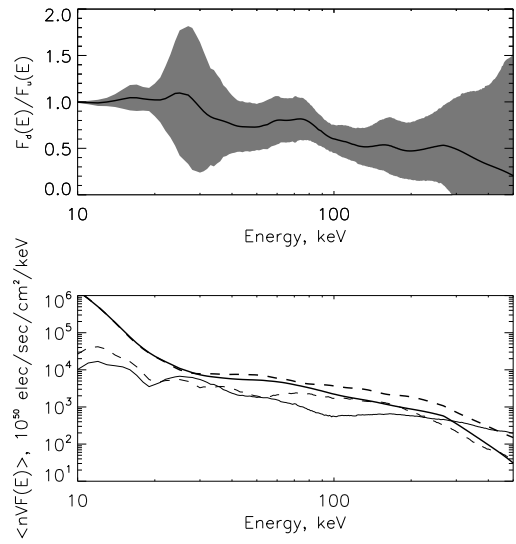}
\end{center}
\caption{Lower panel: The recovered mean electron flux spectra
(thick lines) for August 20, 2002 flare (accumulation time
interval 08:25:20-08:25:40 UT) downward-directed ${\bar F}_d(E)$
(solid line) and observer-directed ${\bar F}_u(E)$ (dash line)
with corresponding errors (thin lines).  Upper panel: electron
anisotropy defined as ${\bar F}_d(E)/{\bar F}_u(E)$ with
confidence values within shaded area.} \label{Aug20}
\end{figure}

To test our method we applied it to simulated data from known
electron distributions, adding typical noise. Two different forms
of anisotropy: `weak' anisotropy ${\bar F}_u(E)= {{\bar
F}_d}(E)/(1+\sqrt{(E-10)/50})$ and `strong' anisotropy ${\bar
F}_u(E)={{\bar F}}_{d}(E)/(1+\sqrt{(E-10)/50})2$ in downward
direction. From the simulated photon spectra we inferred
regularized electron fluxes (${\bar F}_u(E)$,${\bar F}_d(E)$). The
results demonstrate that the method used recovers reliably the
directional electron spectra. However, the method is not sensitive
to too 'weak' anisotropy that gives us anisotropy sensitivity. The
difference between ${\bar F}_u(E)$ and ${\bar F}_d(E)$ for the
case of 'weak' anisotropy is within the error bars.

First we consider some limiting cases.

 {\bf If all electrons were purely downward
directed} (e.g. ${\bar F}_d\neq0$ ${\bar F}_u=0$ in Equation
(\ref{IP})) then the observed spectrum would be mostly given by
reflected and backward emitted photons (Figure 2). Since the
efficiency of Compton scattering as well as the efficiency of
backward emission ($Q_B(E,\epsilon)<<Q_F(E,\epsilon)$ for
$E,\epsilon > 50$ keV) decrease fast with energy, the photon
spectrum should be rather steep. The photon spectral index should
be not less than 5 (Figure 1) in the 200-300 keV energy range even
for a flat power-law electron spectrum while the typical spectral
index observed is around three (Aschwanden, 2002). Moreover, the
inverse approach concludes that for the August 20, 2002 flare
(Kasparova et al, 2005), if the HXRs were strongly downward beamed
(within a pitch angle of $45^o$) the electron spectrum required
would have to grow above 400 keV to fit the data which casts doubt
on the original assumption of downward beaming.

{\bf If we ignore albedo} Equation (\ref{IP}) leads to a
substantial gap or low energy cutoff in the mean electron spectrum
(e.g. Kasparova et al, 2005), hence the local speed distribution
function $f(v)={\bar F}(E)dE/dv/v$ has a positive derivative. Such
distributions are unstable and relax in the solar corona on a
timescale much shorter than our observing time interval (e.g.,
Emslie and Smith 1984, Melnik et al, 1999).

\begin{figure}
\begin{center}
\epsscale{1.0} \plotone{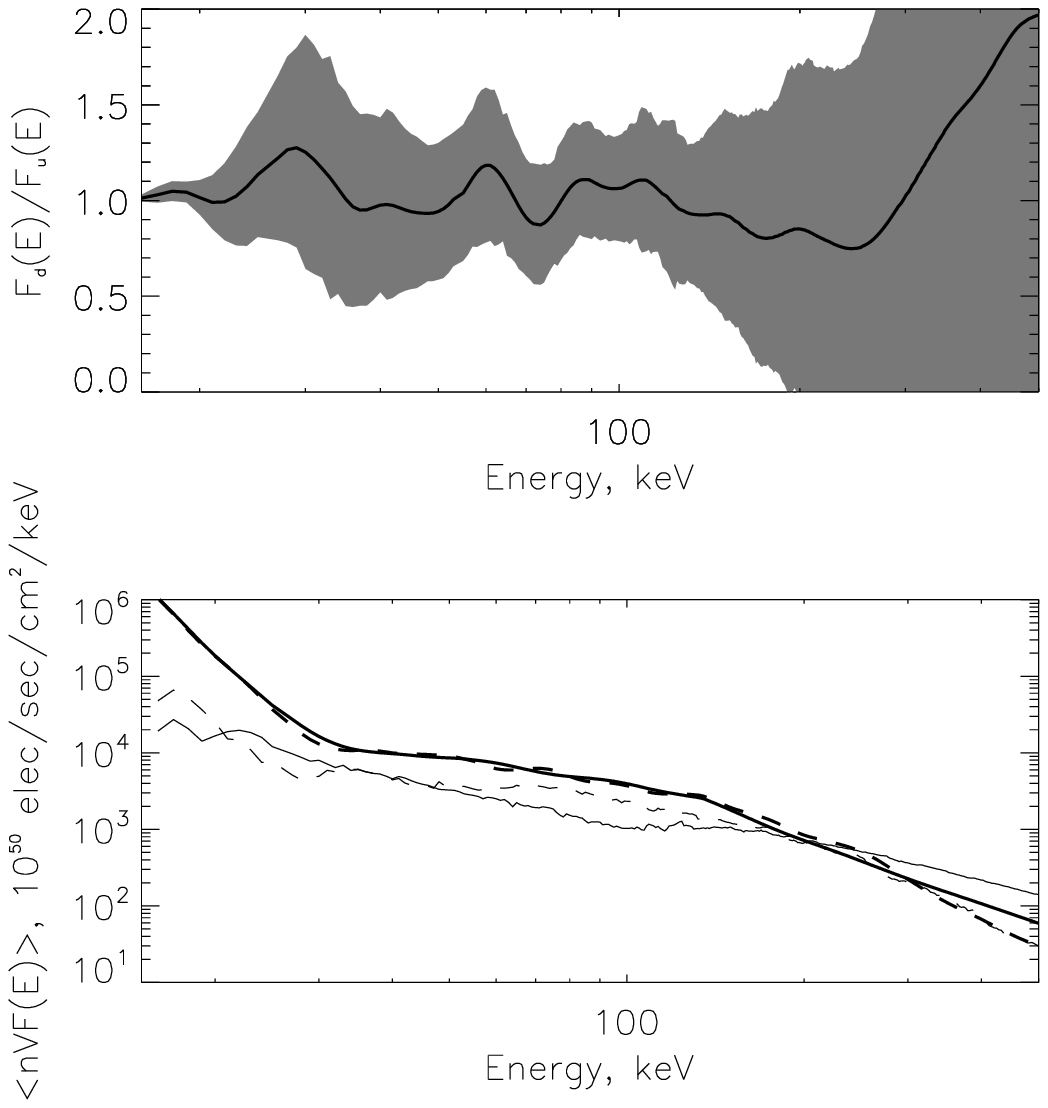}
\end{center}
\caption{The same as Figure (\ref{Aug20}) but for January 17, 2005
flare (accumulation time interval 09:43:24-09:44:20 UT).}
\label{Jan17}
\end{figure}

We now consider the spectra at the peak of two solar flares:
August 20, 2002 around $08:20$~UT (heliocentric angle $\sim 40^o$)
and January 17, 2005 around $09:40$~UT (heliocentric angle $\sim
33^o$). Both have sufficiently high count rates for good
statistics up to above 500 keV but not so high as to cause severe
pulse pile-up effects (Smith et al, 2002). We used 7 out of 9
front RHESSI segments, excluding detectors 2 and 7 due to their
low energy resolution at the time of observation (Smith et al,
2002). Naturally we have chosen events not far from the disk
centre since limb events are not suitable for our analysis because
albedo is almost negligible for such flares.

The general properties of the August 20, 2002 flare $\sim08:20$UT,
which matched our count rate criteria, were extensively analyzed
in Kasparova et al, 2005. To fit the hard X-ray spectrum with a
double power-law electron spectrum with a low energy cut-off $E_c$
and ignoring albedo, requires an unusually high value of $E_c \sim
30\pm2$ keV. This produces a clear gap in the overall ${\bar
F}(E)$ in the range 15-30 keV which is likely to be unphysical and
suggests albedo is important.

Figure \ref{Aug20} shows electron spectra solutions (${\bar
F}_d(E), {\bar F}_u(E)$) with the confidence strips of our
inferred downward and upward electron flux spectra. The results
are close to consistent with isotropy up to 100 keV with some
indication of upward anisotropy above 100 keV. Purely isotropic
distribution would be given by unity, e.g. ${\bar F}_u(E)={\bar
F}_d(E)$. The January 17, 2004 flare has similar characteristics
(Figure \ref{Jan17}): a rather flat spectrum (photon spectral
index $\leq 2.3$) and good count rate up to a few hundred keV.
However, this flare started in the soft X-ray tail of another
event. To avoid difficulties with background substraction, we used
data only from 15 keV.

\section{Discussions and conclusions}

The analysis of RHESSI spectra shows that we can infer
simultaneously information on the directional and the energy
distributions of electrons. It is clear that for the January 17
flare (Figure \ref{Jan17}) that the ${\bar F}_d(E)$, ${\bar
F}_u(E)$ solution are so overlapping and ratio ${\bar
F}_d(E)/{\bar F}_u(E)$ is so close to unity that the electron
distribution is consistent with isotropy at all $E$. For the
August 20 flare in Figure \ref{Aug20}, the distribution is
consistent with isotropy at low $E$ (below 100 keV) but may be
slightly beamed toward the observer at high $E$. Since our
recoveries are means about just 2 directions, a wide range of
actual $\mu$ distributions are consistent with them. For example,
the August 20 event somewhat resembles a 'pancake'  distribution
of particles circling perpendicular to the magnetic field, so that
perpendicular component of electron energy should be slightly
larger than parallel one (Figure \ref{Aug20}). However, none of
these results resembles the beam-like form expected for the basic
CTT model. For example, a simplified mean particle treatment
(Brown, 1972) has no electrons propagating upward, while more
detailed dispersive treatments (MacKinnon and Craig 1991) suggests
downward anisotropies of up to ${\bar F}_d(E)/{\bar F}_u(E)\sim
10-100$ depending on energy and on initial pitch angle
distribution. Full Fokker-Planck models including magnetic
mirroring with a converging magnetic field (Leach and Petrosian
1981) give larger upward flux but still predominantly downward and
outside our confidence interval. For the large fluxes involved,
these events may involve substantial return current $E$-fields
which can produce some upward 'beaming' but the effect is larger
at lower below 70-75~keV, not higher, energies (Zharkova and
Gordovskyy 2005). More advanced self-consistent models with
collective effects included are needed for detailed comparisons
with the observations.

In conclusion, when allowance is made for the albedo contribution
to the observed HXR spectra, the absence of a strong albedo
feature precludes the basic model with strong downward beaming
including CTT models, at least for the two intense hard solar disk
events we have analyzed. This casts serious doubt on the CTT
model, at least in terms of the usually adopted geometry and on
its physical basis. Alternatives to the CTT Standard Model may
need to be considered, in which acceleration occurs in such a way
that collimated injection into a cold target is not involved. For
example, HXR source electrons could be locally and continuously
reaccelerated with near isotropy along the entire length of a
magnetic loop. This model may also solve some of the other
difficulties associated with the CTT model such as the problematic
high beam density involved.



\acknowledgments

We gratefully acknowledge the support of a PPARC Advanced
Fellowship (EPK) and a PPARC Rolling Grant (JCB). We also note
support of ISSI, Bern, Switzerland. We are thankful to Lyndsay
Fletcher, Joe Khan, Alec MacKinnon and S\"am Krucker for comments
and discussions.

\clearpage

\end{document}